\documentclass[prb,amsmath,amsfonts,amssymb,unsortedaddress,floatfix,reprint]{revtex4-1}
\usepackage{dcolumn}
\usepackage{graphicx}
\usepackage{bm}
\usepackage{color}

\begin{document}
\title{Can single-reference coupled cluster theory describe static correlation?}
\author{Ireneusz W. Bulik}
\affiliation{Department of Chemistry, Rice University, Houston, TX 77005-1892}

\author{Thomas M. Henderson}
\affiliation{Department of Chemistry, Rice University, Houston, TX 77005-1892}
\affiliation{Department of Physics and Astronomy, Rice University, Houston, TX 77005-1892}

\author{Gustavo E. Scuseria}
\affiliation{Department of Chemistry, Rice University, Houston, TX 77005-1892}
\affiliation{Department of Physics and Astronomy, Rice University, Houston, TX 77005-1892}
\date{\today}

\begin{abstract}
While restricted single-reference coupled cluster theory truncated to singles and doubles (CCSD) provides very accurate results for weakly correlated systems, it usually fails in the presence of static or strong correlation.  This failure is generally attributed to the qualitative breakdown of the reference, and can accordingly be corrected by using a multi-determinant reference, including higher-body cluster operators in the ansatz, or allowing symmetry breaking in the reference.  None of these solutions are ideal; multi-reference coupled cluster is not black box, including higher-body cluster operators is computationally demanding, and allowing symmetry breaking leads to the loss of good quantum numbers.  It has long been recognized that quasidegeneracies can instead be treated by modifying the coupled cluster ansatz.  The recently introduced pair coupled cluster doubles (pCCD) approach is one such example which avoids catastrophic failures and accurately models strong correlations in a symmetry-adapted framework.  Here we generalize pCCD to a singlet-paired coupled cluster model (CCD0) intermediate between coupled cluster doubles and pCCD, yielding a method that possesses the invariances of the former and much of the stability of the latter.  Moreover, CCD0 retains the full structure of coupled cluster theory, including a fermionic wave function, antisymmetric cluster amplitudes, and well-defined response equations and density matrices.
\end{abstract}
\maketitle

\section{Introduction}
Over the past few decades, single-reference coupled cluster (CC) theory\cite{Paldus1999,Bartlett2007,ShavittBartlett} has reigned unchallenged as the dominant paradigm for the accurate description of weakly correlated systems in quantum chemistry.  This dominance is justly earned.  Coupled cluster delivers very accurate results both for energies and for other properties, with reasonable polynomial scaling.  With an adequate reference determinant, it is size consistent (correct separability) and size extensive (correct scaling with system size).  The method is capable of treating closed shells and open shells, ground states and excited states.  

Where standard forms of single-reference coupled cluster fail is in the description of problems where strong or static correlation effects are important.  In such cases, the failure of single-reference coupled cluster theory is almost always attributed to the reference determinant: the presence of static correlation, we are told, means that the restricted Hartree-Fock (RHF) reference is not even qualitatively reasonable.  This is, of course, true.  But why should the exponential coupled cluster wave function fail catastrophically, as we see, for example, in the dissociation of N$_2$,\cite{Fan2006} or in the large $U$ limit of the Hubbard Hamiltonian,\cite{Hubbard1963} or in the large $G$ limit of the attractive pairing Hamiltonian?\cite{Dukelsky2003,Henderson2014}  To be sure, a correlated wave function built atop a poorly chosen reference need not give accurate predictions, but we would like the predicted energy to at least be real and finite!  In other words, one must tame the failures of coupled cluster theory built atop a symmetry adapted reference.

One way to overcome these failures is simply to work harder.  That is, one can add higher and higher levels of excitation to the cluster operator or at least selected higher excitations; eventually, the cluster operator will contain excitations of sufficiently high rank to overcome the deficiencies of the reference determinant.  After all, coupled cluster theory is formally exact when one includes all possible excitations in the cluster operator.  Multireference methods, in other words, are not needed for strongly correlated systems, if one can use a sufficiently complete single-reference method.  Put this way, the failures of single reference coupled cluster theory for strongly correlated systems are failures of computational ability, not deficiencies of the theory.

The purpose of this manuscript is to emphasize that coupled cluster theory can be protected from these catastrophic failures by working smarter instead.  We seek improvement, that is, by removal rather than by addition.  This is not a new idea.  Numerous authors have proposed methods that yield reasonable predictions in the presence of near degeneracies by modifying coupled cluster theory or variants thereof.\cite{Meyer1971,Jankowski1980,Paldus1984,Piecuch1990,Piecuch1996,VanVoorhis2000,Piecuch2001,Bartlett2006,Neese2009,Huntington2010,Small2012,Kats2013,Limacher2013}  We would like a simple conceptual framework which allows us to identify the offending terms in traditional coupled cluster theory so that by excluding them we can avoid the downfall of the method.  Of course, this is easier said than done, but we here show one particular case which substantially removes the typical failures in repulsive Hamiltonians.

What causes these failures?  Closure of the occupied-virtual gap is commonly cited as the culprit, but this cannot be the whole story.  It is, of course, true that closure of the gap causes \textit{perturbation theory} to fail, but the infinite order summations in coupled cluster theory can overcome a vanishing gap.  For example, the coupled cluster doubles (CCD) equations can schematically be written as
\begin{equation}
\Delta \epsilon \cdot t = v + f(t)
\end{equation}
where $v$ is a two-electron integral; closure of the gap ($\Delta\epsilon \to 0$) just means that we must have $t$ such that $v + f(t) = 0$.  Thus, for example, CCD is well-behaved for the dissociation of H$_2$ despite the gap closing; similarly, we can obtain perfectly reasonable results for the homogeneous electron gas even though it is metallic,\cite{Shepherd2014} or for the repulsive pairing Hamiltonian even when all occupied levels have higher energy than all virtual levels.

If the problem is not the vanishing gap, what else could it be?  The key insight can be obtained by considering the underlying mean-field theory.  As correlations become stronger, RHF becomes unstable toward a symmetry broken unrestricted Hartree-Fock (UHF).  The symmetry spontaneously (but artificially) breaks at the mean-field level, and one can safely build coupled cluster wave functions starting from the UHF reference at the cost of good quantum numbers (symmetry labels) which are difficult to recover once lost.

It seems clear that the failures of single-reference coupled cluster theory are in some way tied to these instabilities.  After all, the particle-hole random phase approximation (RPA) lurks within the coupled cluster equations,\cite{Scuseria2008,Scuseria2013,Peng2013} and the RPA delivers unphysically large correlation energies as RHF approaches the instability point and complex correlation energies past it.  The problems caused by these instabilities are at least partially screened even at the simple CCD level, as is clear by the fact that CCD can dissociate H$_2$ correctly where RPA cannot.  However, fully eliminating the problems caused by these instabilities would appear to require going beyond CCD in general.  This we are loathe to do, simply because it is computationally demanding.  As we have said, we would prefer an alternative framework which does not greatly increase computational complexity.  Much of the inspiration in the present paper comes from the recently introduced pair CCD (pCCD) method,\cite{Limacher2013,Limacher2014,Tecmer2014,Boguslawski2014,Stein2014,Henderson2014b,Henderson2015} which seems immune to problems with instabilities and provides a reasonable description of strong correlation effects.  However, pCCD has its own limitations and requires a full orbital optimization, which is far from being black-box.  Our purpose here is to eliminate as many of the vices of pCCD as possible while retaining most of its virtues.

In the remainder of this paper, we will first illustrate how the failures of CCD can be avoided by removing the terms sensitive to instabilities at the RPA level.  This approach, however, is not ideal, because leaving out entire classes of terms removes the association between the cluster operator on the one hand and an approximate solution to the Schr\"odinger equation on the other.  We remedy this deficiency in Sec. \ref{Sec:CCD0} by removing parts of the cluster operator itself, thereby retaining the eigenvector property, and show several illustrative results before offering concluding remarks.

\section{Failures in Coupled Cluster Doubles}
In this section we wish to provide a few pedagogical examples which illustrate clearly how failures of restricted CCD can be related to symmetry instabilities in the underlying reference determinant and can be ameliorated by simply deleting terms from the amplitude equations.  To do so, we should say a few words about the Hartree-Fock stability conditions.

Solving the Hartree-Fock equations guarantees only that the energy is stationary with respect to orbital rotations.  To obtain a local minimum, we also need that the Hartree-Fock orbital Hessian has only positive eigenvalues.  If the Hessian has a negative eigenvalue, the Hartree-Fock solution is unstable and a lower energy solution can be found.  Particle-hole RPA (ph-RPA) is closely related to the particle-hole Hartree-Fock Hessian, and when this Hessian has a negative eigenvalue, ph-RPA predicts a complex correlation energy.  Similarly, particle-particle (pp-RPA) is related to a Hartree-Fock Hessian which is sensitive toward a number-broken mean-field reference; when the Hartree-Fock is unstable to number symmetry breaking, the Hessian develops a negative eigenvalue and the pp-RPA correlation energy becomes complex.  One can of course use an eigenvector corresponding to a negative eigenvalue to find a mean-field solution with lower energy and typically lower symmetry.  If the mean-field breaks symmetry, the negative eigenvalue disappears in favor of a Goldstone mode with zero eigenvalue in the broken-symmetry Hessian; thus, the RPA built atop a symmetry-broken reference yields a real correlation energy.\cite{BlaizotRipka}  We should emphasize that the RPA we discuss here includes the full exchange interaction.  By neglecting exchange to form direct RPA, one severs the link between RPA and the Hartree-fock stability problem and thereby obtains real correlation energies.\cite{Hesselmann2011,Eshuis2012}

While CCD seems a world removed from stability of Hartree-Fock solutions, contact is made through the RPA.  Both ph-RPA and pp-RPA can be cast as channels of CCD,\cite{Scuseria2008,Scuseria2013,Peng2013} so it seems at least plausible that the Hartree-Fock instabilities which cause the failures of RPA may be ultimately responsible for the failures of CCD as well.  This in turn suggests that these failures might be remedied by deleting the appropriate RPA-like terms from the CCD amplitude equations.

Let us take a moment to be precise.  In CCD, the wave function is given as
\begin{equation}
|\Psi\rangle = \mathrm{e}^{T_2} |0\rangle
\end{equation}
where $|0\rangle$ is a reference determinant (taken here to always be the RHF determinant) and the cluster operator $T_2$ creates double excitations:
\begin{equation}
T_2 = \frac{1}{4} \, \sum_{ijab} t_{ij}^{ab} \, c_a^\dagger \, c_b^\dagger \, c_j \, c_i
\end{equation}
where spinorbitals indexed by $i$, $j$, $k$, $l$ ($a$, $b$, $c$, $d$) are occupied (unoccupied) in $|0\rangle$.  The correlation energy is
\begin{equation}
E = \frac{1}{4} \, \sum_{ijab} t_{ij}^{ab} \, v^{ij}_{ab}
\end{equation}
where $v^{ij}_{ab} = \langle ij \| ab \rangle$ is an antisymmetrized two-electron integral in Dirac notation, while the wave function amplitudes are obtained by solving
\begin{align}
0
 &= f^a_c \, t_{ij}^{cb} + f^b_c \, t_{ij}^{ac} - f_i^k \, t_{kj}^{ab} - f_j^k \, t_{ik}^{ab} + v_{ij}^{ab}
\label{CCDEqns}
\\
 &+ \frac{1}{2} \, t_{kl}^{ab} \, v_{ij}^{kl}
  + \frac{1}{2} \, t_{ij}^{cd} \, v_{cd}^{ab}
  + \frac{1}{4} \, t_{ij}^{cd} \, t_{kl}^{ab} \, v_{cd}^{kl}
\nonumber
\\
 &+ t_{ik}^{ac} \, v^{kb}_{cj}
  + t_{kj}^{cb} \, v^{ka}_{ci}
  + t_{ik}^{ac} \, t_{jl}^{bd} \, v^{kl}_{cd}
\nonumber
\\
 &- t_{ik}^{bc} \, v^{ka}_{cj}
  - t_{kj}^{ca} \, v^{kb}_{ci}
  - t_{ik}^{bc} \, t_{jl}^{ad} \, v^{kl}_{cd}
\nonumber
\\
 &- \frac{1}{2} \, v^{kl}_{cd} \, \left(t_{il}^{cd} \, t_{kj}^{ab} + t_{lj}^{cd} \, t_{ik}^{ab} + t_{kl}^{ad} \, t_{ij}^{cb} + t_{kl}^{db} \, t_{ij}^{ac}\right)
\nonumber
\end{align}
in terms of the Fock operator $f^p_q = \langle q | f | p \rangle$; in this equation, summation over repeated indices is implied.  We refer to the terms on the first line of the CCD amplitude equation as the driver; the terms on the next three lines we call the ladder, ring, and crossed-ring terms, and we call the terms on the last line mosaic terms.\cite{Scuseria2013}  One can obtain ph-RPA from CCD simply by retaining only the driver and ring terms (modulo an overall factor of two on the correlation energy which in this work we will neglect), and pp-RPA by retaining only the driver and ladder terms.

All of this means that if the RPA instabilities drive failures of CCD, one should obtain well-behaved results by eliminating the ladder terms in systems for which pp-RPA exhibits instabilities and by eliminating the ring terms in systems for which ph-RPA becomes unstable; in the latter case, one should also remove the crossed-ring terms so as to retain proper fermionic antisymmetry of the amplitudes $t_{ij}^{ab}$.  To see that eliminating the offending RPA terms does indeed cure the failures of CCD, we provide several illustrative examples.

\begin{figure}[t]
\includegraphics[width=0.5\textwidth]{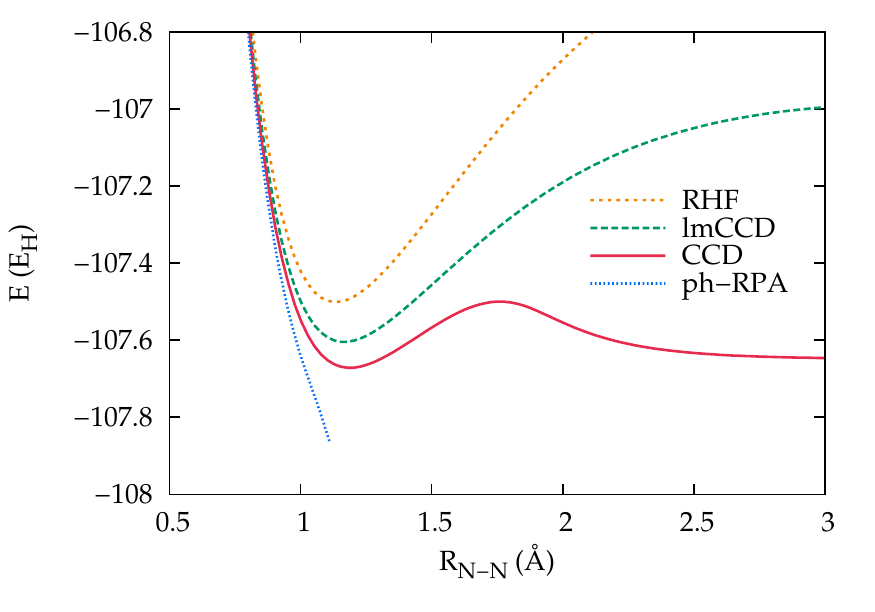}
\caption{Total energies of the N$_2$ molecule in the STO-3G basis set as a function of bond length.  Past the point shown, ph-RPA yields a complex correlation energy.  ``lm-CCD'' denotes CCD where we have eliminated ring and crossed-ring terms.
\label{Fig:N2Fail}}
\end{figure}

Consider, then, the dissociation of the N$_2$ molecule.  We will work in the STO-3G basis,\cite{STO3G} which is useless for coupled cluster calculations in general but which here serves to isolate the effects of static correlation fairly well as we have essentially a valence-only calculation.  As Fig. \ref{Fig:N2Fail} shows, CCD breaks down very badly in this case, predicting a dissociation limit very close to the equilibrium energy.  The ph-RPA energy is disastrous everywhere, and becomes complex near equilibrium.  On the other hand, removing ring and crossed-ring contractions to obtain what we have called lmCCD\cite{Footnote1}
produces a well-behaved dissociation curve (in that the energy rises monotonically from equilibrium to dissociation), albeit one which is by no means energetically satisfactory.

\begin{figure}[t]
\includegraphics[width=0.5\textwidth]{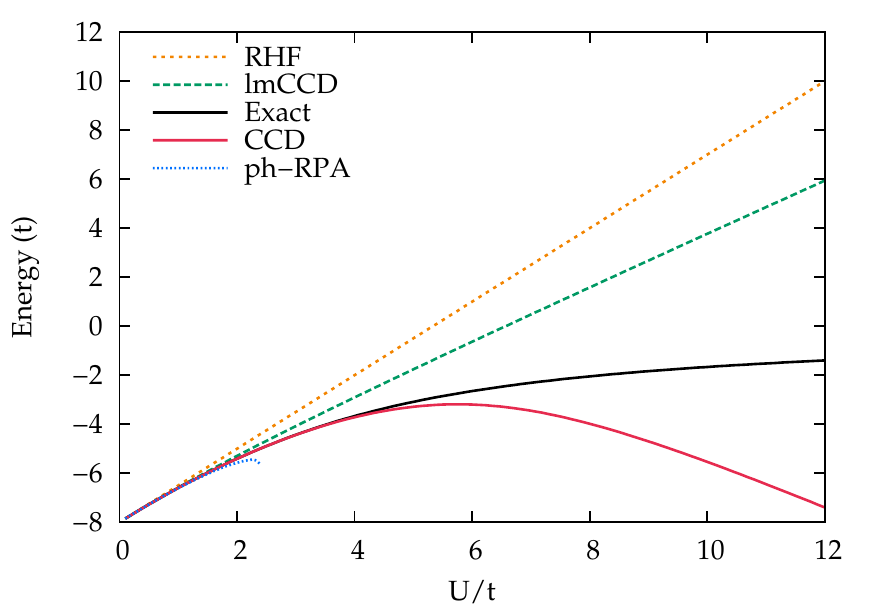}
\caption{Total energies in the 6-site Hubbard Hamiltonian at half filling as a function of interaction strength $U/t$.  The ph-RPA yields a complex correlation energy past the point shown.  ``lm-CCD'' denotes CCD where ring and crossed-ring terms have been excluded.
\label{Fig:HubbardFail}}
\end{figure}

The same happens in the repulsive Hubbard Hamiltonian, which is given by
\begin{equation}
H = -t \sum_{\langle pq \rangle, \sigma} c_{p_\sigma}^\dagger \, c_{q_\sigma} + U \, \sum_p n_{p_\uparrow} \, n_{p_\downarrow}
\end{equation}
where $p$ and $q$ index lattice sites and $\sigma$ indexes spins, $n_{p_\sigma} = c_{p_\sigma}^\dagger \, c_{p_\sigma}$ is a spinorbital number operator, and the notation $\langle pq \rangle$ indicates that sites $p$ and $q$ are nearest neighbors; we take the lattice to be periodic so that sites 1 and $N$ in an $N$-site Hubbard Hamiltonian are adjacent.  In Fig. \ref{Fig:HubbardFail}, we see that CCD fails again for the 6-site Hubbard model at half-filling.  While the exact energy rises monotonically to zero as $U$ approaches infinity, the CCD energy quickly turns over and gets increasingly negative as $U$ gets larger, and ph-RPA yields a complex correlation energy even for relatively small $U$.  Again, removing the ring and crossed-ring terms cures this qualitative failure without, however, providing reasonable energies.

\begin{figure}[t]
\includegraphics[width=0.5\textwidth]{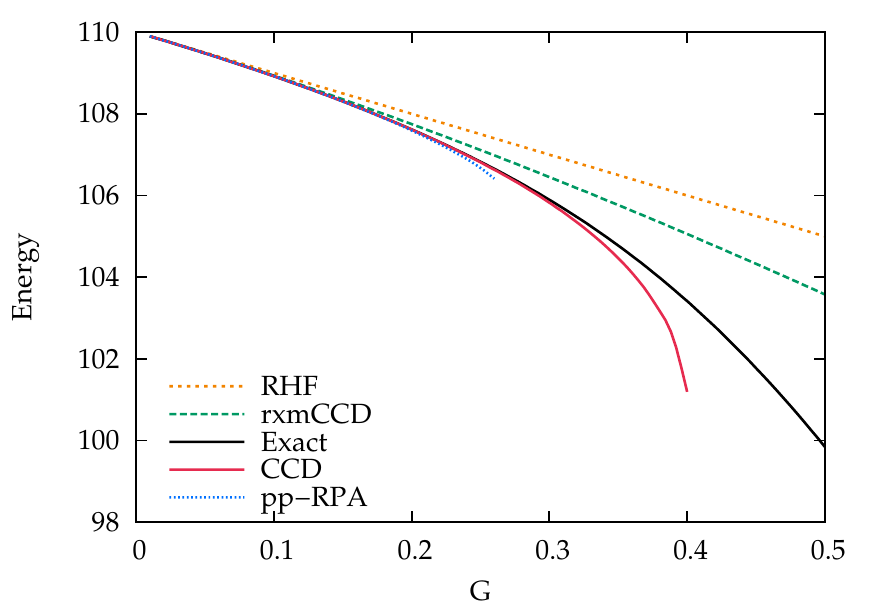}
\caption{Total energies in the attractive pairing Hamiltonian as a function of interaction strength $G$.  Past the points shown, pp-RPA and CCD yield a complex correlation energy.  ``rxm-CCD'' denotes CCD where laddter terms have been omitted.
\label{Fig:PairingFail}}
\end{figure}

Lastly, consider the attractive pairing Hamiltonian, given by
\begin{equation}
H = \sum_{p,\sigma} \epsilon_p \, n_{p_\sigma} - G \, \sum_{pq} c_{p_\uparrow}^\dagger \, c_{p_\downarrow}^\dagger \, c_{q_\downarrow} \, c_{q_\uparrow},
\end{equation}
where $G > 0$ and the levels are equally spaced ($\epsilon_p = p$).  For this problem, there is an instability toward number symmetry breaking.  Both pp-RPA and in fact CCD eventually predict complex correlation energies, but as can be seen from Fig. \ref{Fig:PairingFail}, removing the ladder terms in what we have denoted rxmCCD provides a stable but inaccurate curve.

It is clear that we can indeed cure the failures of CCD by simply deleting entire classes of terms from the amplitude equations.  The cure, however, is often worse than the disease.  We shall do better shortly, but it is instructive to see why eliminating terms from the CCD equations may yield poor results.

The CCD energy and amplitude equations shown earlier are given by
\begin{subequations}
\begin{align}
E_\mathrm{CCD} &= \langle 0 | \mathrm{e}^{-T_2} \, H \, \mathrm{e}^{T_2} | 0\rangle,
\\
0 &= \langle \Phi_{ij}^{ab} | \mathrm{e}^{-T_2} \, H \, \mathrm{e}^{T_2} | 0\rangle,
\end{align}
\end{subequations}
where the state $|\Phi_{ij}^{ab} \rangle$ is a doubly-excited determinant.  It is clear that CCD can be viewed as writing a similarity-transformed Hamiltonian $\bar{H} = \exp(-T_2) \, H \, \exp(T_2)$ such that the reference determinant $|0\rangle$ is a right-hand eigenvector within the space of the reference determinant and all double excitations.  In other words, the CCD amplitude equations are obtained by solving the Schr\"odinger equation for a transformed Hamiltonian within a subspace.  This is no longer the case when one removes terms whole cloth.  In other words, when one eliminates entire classes of terms from the CCD amplitude equations, one is no longer solving the Schr\"odinger equation; accordingly, it should not be too surprising that the results thereby obtained are unsatisfactory.

\begin{figure*}[t]
\includegraphics[width=0.45\textwidth]{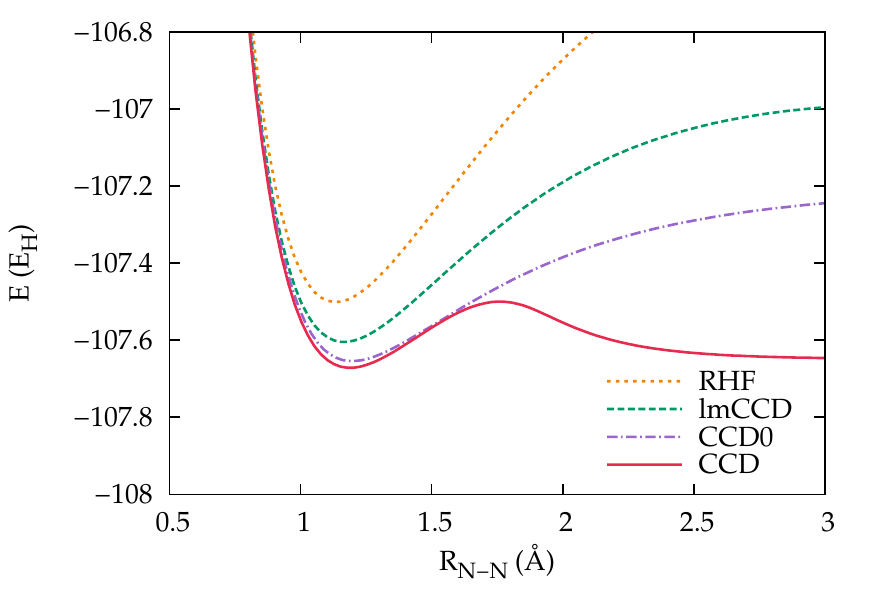}
\includegraphics[width=0.45\textwidth]{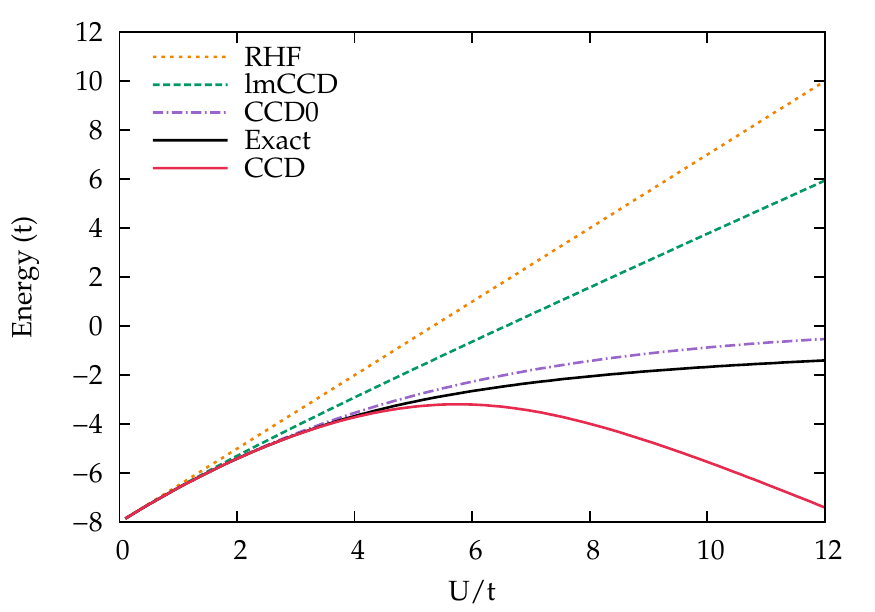}
\caption{Left panel: Total energies of the N$_2$ molecule in the STO-3G basis set as a function of bond length.  Right-panel: Total energies in the 6-site Hubbard Hamiltonian at half filling as a function of interaction strength $U/t$.  In both cases, ``lm-CCD'' denotes CCD where particle-hole contractions have been excluded.
\label{Fig:CCD0Intro}}
\end{figure*}

\section{Singlet-Paired CCD
\label{Sec:CCD0}}
Ideally, we would like to formulate CCD in terms of a cluster operator which is insensitive to instabilities in the reference without having to remove entire classes of terms, because by doing so, we guarantee that we have a solution to the Schr\"odinger equation at least within a subspace of the full Hilbert space of the problem.  While it is not entirely clear how one might accomplish this in a general way, we do have a specific example of a CCD model which avoids these problems.  That model is pCCD, in which the cluster operator is rewritten as simply
\begin{equation}
T_2^\mathrm{pCCD} = \sum_{ia} T_i^a \, c_{a_\uparrow}^\dagger \, c_{a_\downarrow}^\dagger \, c_{i_\downarrow} \, c_{i_\uparrow}
\end{equation}
where here and for the remainder of this manuscript orbital labels $i$, $j$, $a$, $b$, and so forth refer to spatial orbitals.  Thus, for example, $c_{a_\uparrow}^\dagger \, c_{a_\downarrow}^\dagger \, c_{i_\downarrow} \, c_{i_\uparrow}$ removes two electrons from spatial orbital $i$ and places them in spatial orbital $a$.  Remarkably, this simplification leads to a well-behaved model which neither overcorrelates wildly nor predicts a complex correlation energy in systems which are unstable toward a UHF reference.  To some extent, we can attribute the success of pCCD in this regard to the fact that it can include the perfect pairing valence bond and antisymmetrized product of strongly orthogonal geminals wave functions, albeit with coefficients determined by similarity transformation rather than via the varitional principle.\cite{Limacher2013}

However, while restricted pCCD provides a reasonable accounting for the paired correlations, it does not include any correlations between electrons in different spatial orbitals.  Moreover, the result of a pCCD calculation is not invariant to rotations amongst the occupied orbitals or amongst the virtual orbitals, which necessitates an orbital optimization which can be somewhat tedious.  The question then becomes whether we can eliminate these flaws without reintroducing the failures for strongly correlated systems.

In order to do so, however, it will prove useful to first revisit standard CCD in the RHF framework, for which the cluster operator is singlet spin adapted and can be conveniently written as\cite{Scuseria1988}
\begin{equation}
T _2= \frac{1}{2} \, \sum_{ijab} T_{ij}^{ab} \, \sum_{\sigma\xi} c_{a_\sigma}^\dagger \, c_{b_\xi}^\dagger \, c_{j_\xi} \, c_{i_\sigma}.
\end{equation}
It is important to note that as a two-body operator, $T_2$ actually contains two independent singlet components which we have parameterized by the single set of amplitudes $T_{ij}^{ab} = T_{ji}^{ba} = t_{i_\uparrow j_\downarrow}^{a_\uparrow b_\downarrow}$.\cite{Geertsen1986,Scuseria1988,Piecuch1990}  That is, we have
\begin{subequations}
\begin{align}
T_2 &= T_2^{[0]} + T_2^{[1]},
\\
T_2^{[0]} &= \frac{1}{2} \, \sum_{ijab} \sigma_{ij}^{ab} \, P_{ab}^\dagger \, P_{ij}
\\
T_2^{[1]} &= \frac{1}{2} \, \sum_{ijab} \pi_{ij}^{ab} \, \vec{Q}_{ab}^\dagger \, \cdot \vec{Q}_{ij}
\end{align}
\end{subequations}
where the pair operators $P_{ab}$ and $\vec{Q}_{ab}$ are given by
\begin{subequations}
\begin{align}
P_{ij} &= \frac{1}{\sqrt{2}} \, \left(c_{j_\uparrow} \, c_{i_\downarrow} + c_{i_\uparrow} \, c_{j_\downarrow}\right),
\\
Q_{ij}^+ &= c_{j_\uparrow} \, c_{i_\uparrow},
\\
Q_{ij}^- &= c_{j_\downarrow} \, c_{i_\downarrow},
\\
Q_{ij}^0 &= \frac{1}{\sqrt{2}} \, \left(c_{j_\uparrow} \, c_{i_\downarrow} - c_{i_\uparrow} \, c_{j_\downarrow}\right)
\end{align}
\end{subequations}
and where the notation $\vec{Q}_{ab}^\dagger \cdot \vec{Q}_{ij}$ is shorthand for 
\begin{equation}
\vec{Q}_{ab}^\dagger \, \cdot \vec{Q}_{ij} = \left(Q_{ab}^+\right)^\dagger \, Q_{ij}^+ + \left(Q_{ab}^-\right)^\dagger \, Q_{ij}^- + \left(Q_{ab}^0\right)^\dagger \, Q_{ij}^0.
\end{equation}
Notice that $P_{ij}$ is symmetric on interchange of $i$ with $j$ whereas $\vec{Q}_{ij}$ is antisymmetric; thus, $\sigma_{ij}^{ab}$ and $\pi_{ij}^{ab}$ obey respectively
\begin{subequations}
\begin{align}
\sigma_{ij}^{ab} &= \sigma_{ij}^{ba} = \sigma_{ji}^{ab} = \sigma_{ji}^{ba} = \frac{T_{ij}^{ab} + T_{ij}^{ba}}{2},
\\
\pi_{ij}^{ab} &= -\pi_{ij}^{ba} = -\pi_{ji}^{ab} = \pi_{ji}^{ba} = \frac{T_{ij}^{ab} - T_{ij}^{ba}}{2}
\end{align}
\end{subequations}
so that $\sigma_{ij}^{ab}$ ($\pi_{ij}^{ab}$) is the part of $T_{ij}^{ab}$ symmetric (antisymmetric) on interchange of $i$ with $j$.  The operators $P_{ij}$ are singlet-paired, while $\vec{Q}_{ij}$ are triplet paired; in other words, the wave function created by $P^\dagger |-\rangle$ is a singlet, while those created by the three components of $Q^\dagger |-\rangle$ are triplets, where $|-\rangle$ is the physical vacuum.  Note that $T_2^{[0]}$ correlates only electrons of opposite spin, while $T_2^{[1]}$ includes both opposite-spin and same-spin correlation.  However, both $T_2^{[0]}$ and $T_2^{[1]}$ yield contributions from all channels of CCD, \textit{i.e.} they retain ladder, ring, crossed-ring, and mosaic terms.

Clearly, pCCD eliminates the triplet-paired operators entirely and retains only the diagonal $i=j$ and $a=b$ portions of the singlet-paired operators.  In other words, in pCCD we eliminate $T_2^{[1]}$ entirely and retain only the diagonal portion of $T_2^{[0]}$.  We could add correlations between electrons in different spatial orbitals and restore invariance to occupied-occupied and virtual-virtual mixing by simply relaxing the restriction that we include only the diagonal portions of the singlet-paired operators.  Put differently, we can generalize pCCD to what we will call CCD0 by simply writing
\begin{equation}
T_2^\mathrm{CCD0} = T_2^{[0]}.
\end{equation}
Operationally, this just means that in a standard RHF-based CCD code which uses $t_{i_\uparrow j_\downarrow}^{a_\uparrow b_\downarrow}$ as the fundamental ingredient, as many codes do,\cite{Scuseria1987} one needs only to symmetrize the amplitudes $T_{ij}^{ab}$ iteration by iteration; in other words, one makes the replacement $T_{ij}^{ab} \to 1/2 \, \left(T_{ij}^{ab} + T_{ij}^{ba}\right)$ every iterative cycle.

Figure \ref{Fig:CCD0Intro} shows the results of CCD0 for the dissociation of N$_2$ and for the Hubbard Hamiltonian.  It is clear that CCD0 remedies the failures of CCD while still including much of the correlations that are lost by eliminating the ring terms in CCD.  These examples, however, are fairly trivial, so we will present several more results to establish the point and to show how one can build atop CCD0 in a natural way.

\begin{figure}[t]
\includegraphics[width=0.35\textwidth]{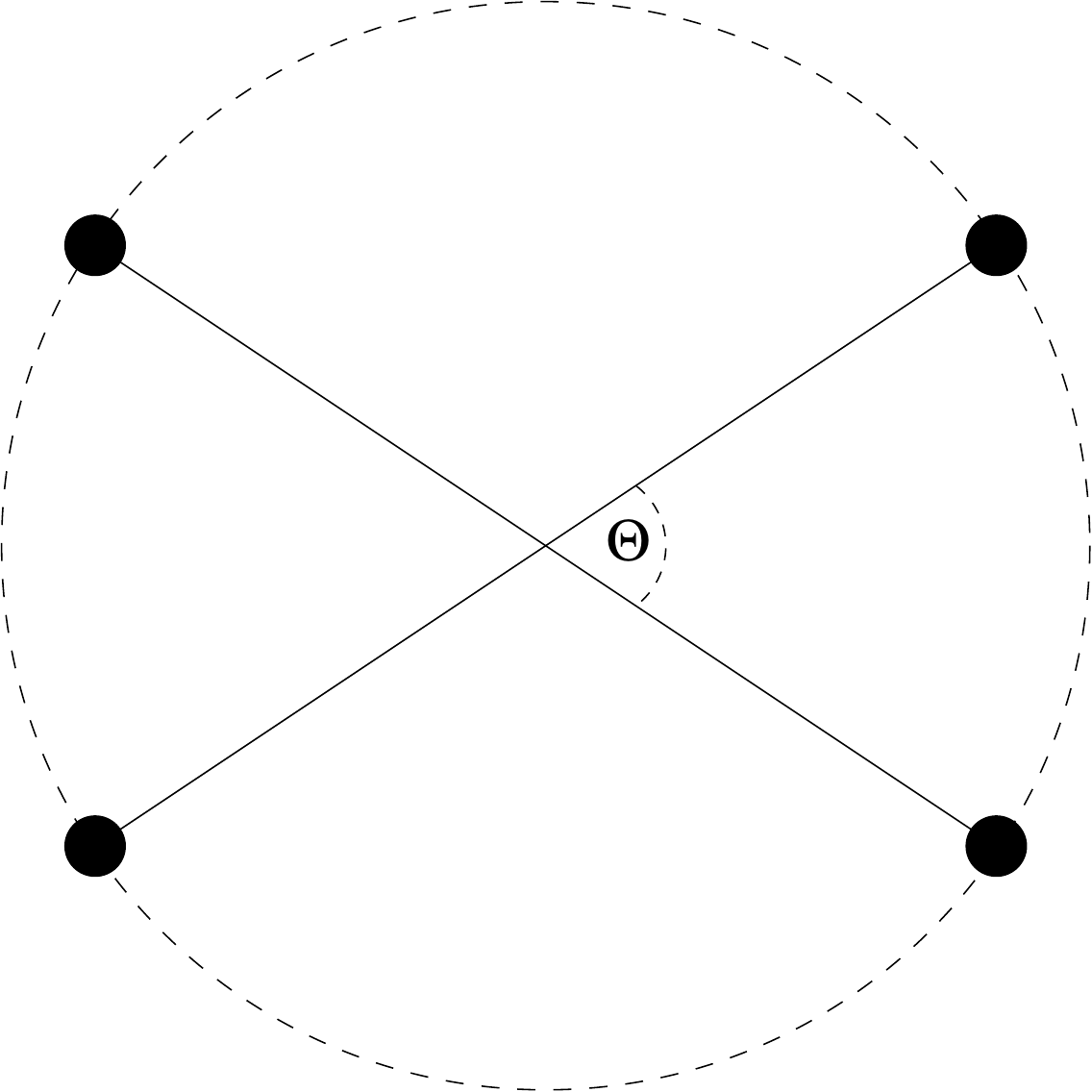}
\caption{Geometry of H$_4$ on a circle.
\label{Fig:H4Circle}}
\end{figure}

\begin{figure}[t]
\includegraphics[width=0.45\textwidth]{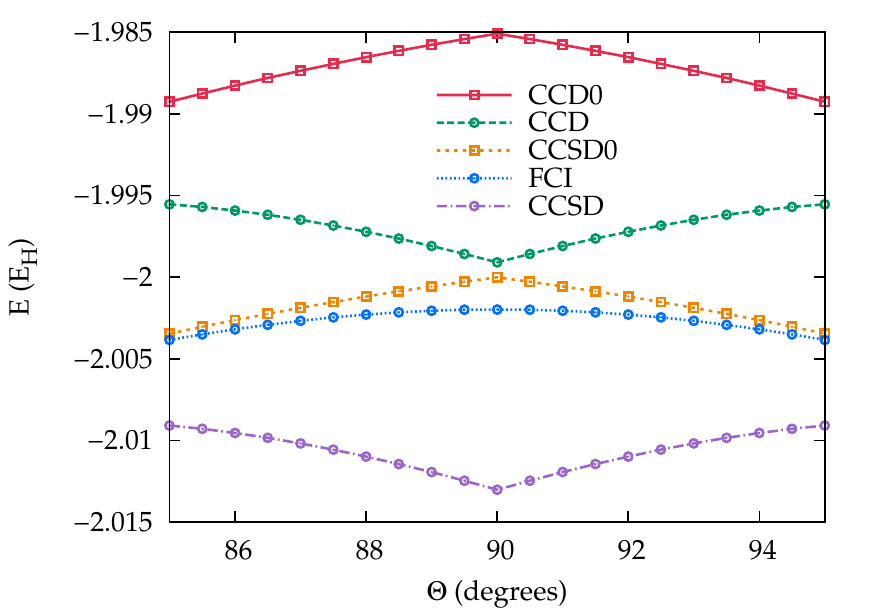}
\caption{Total energies of H$_4$ as a function of angle $\Theta$.
\label{Fig:H4Results}}
\end{figure}

A strict test for coupled cluster is the case of four hydrogen atoms symmetrically distributed on a circle of radius $R = 1.738 \mathrm{\AA}$, as depicted in Fig. \ref{Fig:H4Circle}.\cite{Troy2000}  For small or large angles $\Theta$, the system resembles two H$_2$ units that are reasonably well separated, but as $\Theta$ passes through $90^\circ$, the four atoms form a square and there is a degeneracy.  The exact energy is smooth as a function of $\Theta$, but at the RHF level there is a cusp at $90^\circ$, just as with the rotation of the C-C bond in ethylene.  Figure \ref{Fig:H4Results} shows total energies for H$_4$ in Dunning's DZP basis set,\cite{DunningDZP,DunningDZP2} all built atop the RHF reference.  Both CCD and coupled cluster with singles and doubles (CCSD) have a cusp and incorrectly predict a local minimum at 90$^\circ$.  While the cusp remains in CCD0 and in CCSD0 (\textit{i.e.} CCD0 with the full inclusion of single excitations), it is smoothed out somewhat and both CCD0 and CCSD0 correctly predict a maximum in the energy at $90^\circ$ instead.  Qualitatively, then, CCD0 and CCSD0 are much closer to correct than are CCD and CCSD; CCSD0 is fortuitously semi-quantitative for this problem but not, as we shall see, in general.

\begin{figure}[t]
\includegraphics[width=0.45\textwidth]{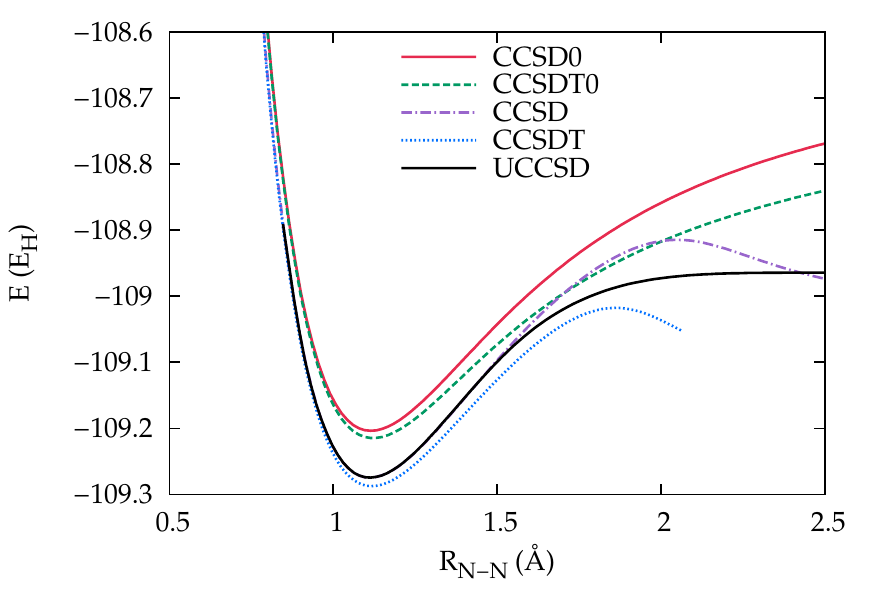}
\caption{Total energies of the N$_2$ molecule in the cc-pVDZ basis set as a function of bond length.  Where CCSD and CCSDT break down for large $R$, CCSD0 and CCSDT0 are properly stable.
\label{Fig:N2Triples}}
\end{figure}

Next, we consider N$_2$ with the slightly more complete cc-pVDZ basis set,\cite{Dunning1989} as seen in Fig. \ref{Fig:N2Triples}.  The failure of CCD (and therefore of CCSD) is lessened in the larger basis, which we can qualitatively understand as arising from the much larger number of virtual orbitals allowing for significantly more dynamic correlation.  Nonetheless, CCSD breaks down in the larger basis where CCSD0 does not.  One could also employ Brueckner orbitals in a CCD0 analogue of Brueckner doubles, which we would denote as BD0, but results do not differ much from those of CCSD0, so we have not shown them here.

These calculations permit us also to make a second point.  In traditional coupled cluster theory, only the $T_1$ and $T_2$ equations are nonlinear in their respective operators, \textit{i.e.} the $T_3$ equations are linear in $T_3$ (though they contain terms like $H \, T_2 \, T_3$), the $T_4$ equations are linear in $T_4$, and so on.  If the failures in CCD can be attributed to the terms quadratic in $T_2$, therefore, it seems reasonable to suppose that including higher-order cluster operators atop CCD0 would immediately lead to a well-behaved method.  This is indeed the case.  Figure \ref{Fig:N2Triples} also shows results with coupled cluster with single, double, and triple excitations (CCSDT),\cite{Noga,GusCCSDT} as well as with what we call CCSDT0, by which we mean CCSDT subject to the restriction that $T_2$ be singlet paired as in CCD0.  Clearly, CCSDT has the same qualitative failures as do CCD and CCSD, but once the appropriate modifications are made to the $T_2$ operator, the resulting method is again well-behaved.  Adding correlations atop CCD0, in other words, is fairly straightforward.  The only challenge is appropriately correcting for the triplet-paired portion $T_2^{[1]}$ of $T_2$ which CCD0 has omitted; this portion is clearly important for getting accurate results in regions where restricted CCD works well, but must be included carefully so as not to reproduce the failures of CCD as well as its successes.

\subsection{Instabilities in CCD0}
Thus far, we have shown no examples for which CCD0 breaks down.  In fact, while CCD0 is much less prone to failure for strongly correlated systems than is CCD, it is not impervious to these problems.  When problems are sufficiently strongly correlated, that is to say, CCD0 is a great improvement upon CCD but is not necessarily a panacea.

\begin{figure}[t]
\includegraphics[width=0.5\textwidth]{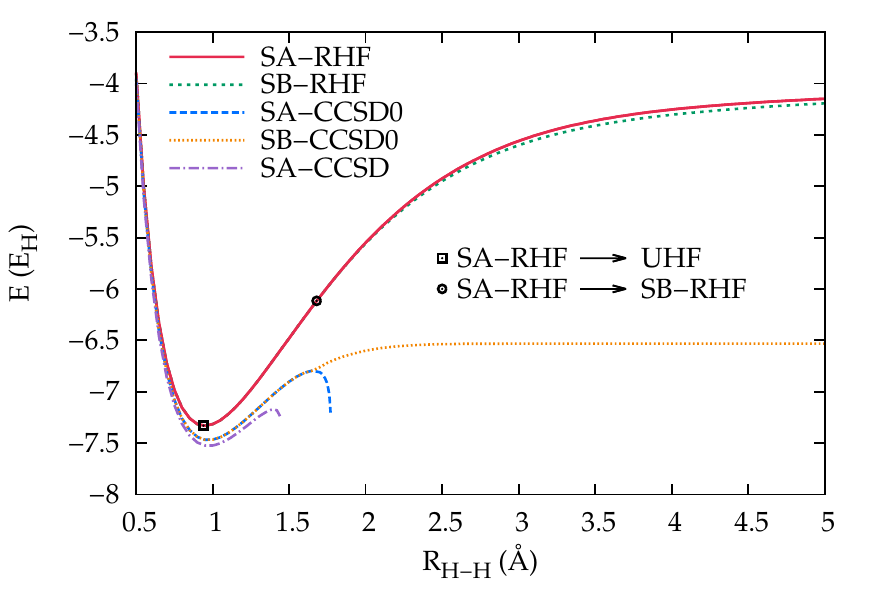}
\caption{Total energies for the H$_{14}$ ring as a function of interatomic separation.  The curves marked ``SA-RHF'' and ``SB-RHF'' are RHF solutions which respectively preserve or break spatial symmetry, while ``SA-CCSD0'' and ``SB-CCSD0'' are CCSD0 curves based on the SA-RHF and SB-RHF references.  The SA-CCSD curve is CCSD based on the SA-RHF reference.  We could not converge CCSD based upon the SB-RHF determinant.  The black points show the onset of instability of the SA-RHF determinant toward UHF or the SB-RHF state.
\label{Fig:H14}}
\end{figure}

Figure \ref{Fig:H14} shows results for a ring of fourteen equally-spaced hydrogen atoms in the STO-3G basis set.  As the radius of the ring increases, the spacing between adjacent atoms increases as well.  Loosely, this forms an analogy with the Hubbard Hamiltonian, though in the Hubbard Hamiltonian the electron-electron interaction is local and has zero range.  Even before equilibrium, there is an instability in the symmetry-adapted RHF (marked as ``SA-RHF'' on the plot) toward a UHF solution.  For larger bond lengths, the symmetry-adapted RHF develops an additional instability, toward another RHF solution which breaks spatial symmetry (marked as ``SB-RHF'') on the plot.  This symmetry-broken RHF has a dimer-like structure, and the occupied orbitals essentially resemble seven sets of H$_2$ bonding orbitals.

Conventional CCSD based on the symmetry-adapted RHF reference begins to overcorrelate dramatically shortly past equilibrium, after which we can no longer converge the CCSD equations.  We were unable to converge the CCSD equations on the symmetry-broken RHF reference.  In contrast, we were able to converge CCSD0 based on the symmetry-adapted reference until much larger bond lengths.  At the point for which the symmetry-adapted RHF develops an instability toward spatial symmetry breaking, the CCSD0 begins to break down.  We can, however, solve this problem by using the symmetry-broken RHF as a reference for CCSD0.  The same general observations hold for CCD and CCD0.

\begin{figure}[t]
\includegraphics[width=0.5\textwidth]{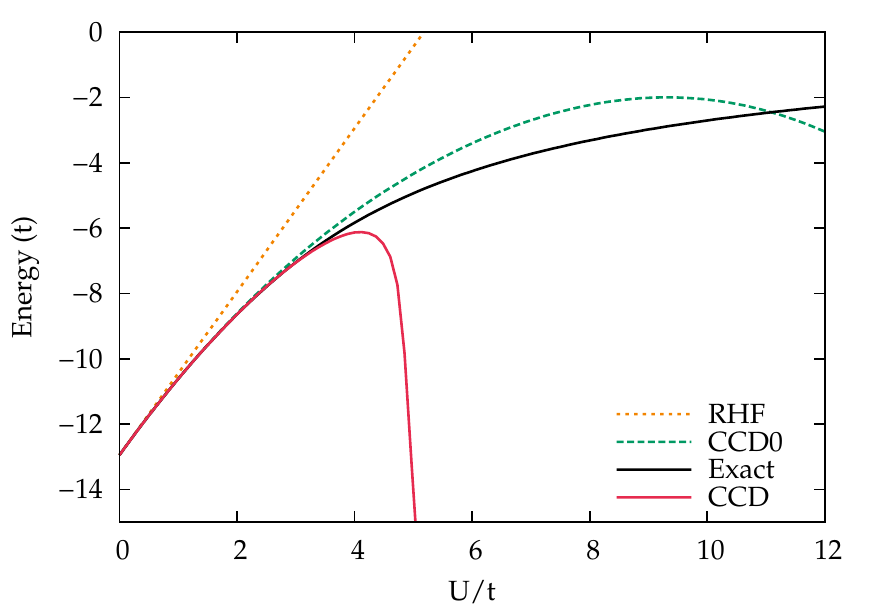}
\caption{Total energies in the 10-site Hubbard Hamiltonian at half filling as a function of interaction strength $U/t$.  
\label{Fig:Hubbard10}}
\end{figure}

Because this hydrogen ring is loosely analogous to the Hubbard Hamiltonian, it may not be too surprising that the Hubbard Hamiltonian displays a similar phenomenon.  In Fig. \ref{Fig:Hubbard10} we show results for the 10-site Hubbard Hamiltonian.  We see that CCD fails dramatically and CCD0 begins to break down for large enough $U/t$.  This is, in fact, also true for the smaller 6-site Hubbard problem, where the CCD energy reaches a maximum near $U/t \approx 6.2$ and the CCD0 energy is maximal for $U/t \approx 20$; eliminating the ring and crossed-ring contractions entirely, in contrast, undercorrelates badly but leaves a total energy which increases monotonically at least until $U/t = 30$.  We have not yet been able to explain the origin of the failure of CCD0 for this problem although such large $U$ values are normally outside the regime of physical systems.  The RHF solution we have used as a reference is stable toward spatial symmetry breaking, unlike with the hydrogen ring, so it is not clear how to use a different RHF reference to resolve this problem.  We have traced the failure of CCD0 in this case to the ring diagrams, as removing the ring and crossed-ring terms leaves, as one might expect, a well-behaved but inaccurate result.

\begin{figure}[t]
\includegraphics[width=0.5\textwidth]{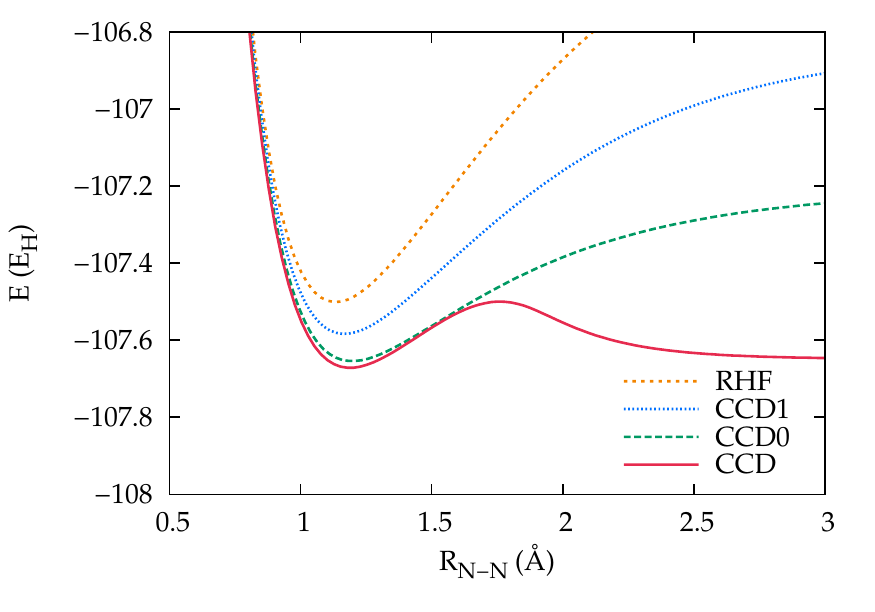}
\caption{Total energies of the N$_2$ molecule in the STO-3G basis set as a function of bond length.  
\label{Fig:N2CCD1}}
\end{figure}

\subsection{Triplet-Paired CCD}
Thus far, we have shown that by simply choosing the pair creation and annihilation operators defining the $T_2$ cluster operator to be singlet paired only, we obtain a method which is largely free of the major failures of CCD.  One could alternatively have chosen the pair creation and annihilation operators to be triplet paired.  In other words, instead of retaining only $T_2^{[0]}$ we could have retained instead only $T_2^{[1]}$.  This defines what we call CCD1, where the ``1'' indicates the $S=1$ triplet-pairing nature of the correlator.  Where CCD0 entirely neglects correlations between same-spin electrons, such correlations are present in CCD1, though at the price of less focus on the more dominant opposite-spin correlations.  We emphasize again that the classification of correlations into ``same spin'' and ``opposite spin'' is different from the classification into singlet- and triplet-paired correlations.  Singlet pairing is opposite-spin only but does not contain all opposite-spin correlations; the remainder of the opposite-spin correlations, together with all same-spin correlations, are contained within the triplet pairing channel.  In Fig. \ref{Fig:N2CCD1} we show the dissociation of N$_2$ in the minimal basis one more time, now including CCD1.  As one can see, CCD1, like CCD0, is generally stable and undercorrelating.  This is also clearly true in the Hubbard Hamiltonian, for which CCD1 predicts zero correlation energy (because the two-electron interaction in the Hubbard Hamiltonian is singlet paired only).  We emphasize that RHF-based CCD has a triplet pairing channel for which the three distinct components all have the same wave function amplitudes.  In UHF-based CCD, one could attempt to define a triplet pairing channel but due to broken spin symmetry would obtain three distinct sets of $T_2$ amplitudes.

Individually, CCD0 and CCD1 are generally well-behaved and undercorrelating.  It is apparent that one cannot meaningfully just solve the CCD0 equations for the symmetric part of the cluster amplitude $T_{ij}^{ab}$ and the CCD1 equations for the antisymmetric part; in the Hubbard Hamiltonian, this would return the CCD0 result, while in N$_2$ it would overcorrelate wildly.  The coupling between these singlet- and triplet-paired parts of $T_2$ which is present in the full CCD equations thus causes CCD to deliver reasonable energies in weakly correlated systems; however, the interaction between singlet- and triplet-paired channels also apparently causes the failures of CCD for strongly correlated systems.  It seems at least possible, then, that modifying the coupling between these two parts of $T_2$ might provide a useful alternative, albeit one unexplored here.  It is evident that full inclusion of higher cluster amplitudes ($T_3$, $T_4$, and so on) serves to renormalize the coupling between the $T_2$ amplitudes.

\section{Conclusions
\label{Sec:Conclusions}}
The failure of traditional coupled cluster theory in strongly correlated systems is well known, and is usually understood as a failure of the reference determinant to even qualitatively describe the wave function.  When such failures take place, they can be ameliorated in one of two ways: either the reference must be improved, or the correlator used to create the correlated wave function from the reference must be improved.  Neither option, unfortunately, is entirely ideal.  Improving the reference can be done by adopting some form of multireference coupled cluster approach, but this is by no means straightforward.  Improving the correlator by making it more complete -- that it, by approaching nearer to full configuration interaction -- is certainly conceptually straightforward, but computationally can be very demanding.  Typically we perforce resort to the use of a broken symmetry mean-field reference; for repulsive two-body interactions, spin symmetry is broken, while for attractive two-body interactions, we break number symmetry instead.\cite{Henderson2014}

What has been explored less is improving the correlator by \textit{removing} pieces.  This has the advantage of computational simplicity but the disadvantage that what is to be removed must be chosen with a certain amount of care.  The point of this manuscript is simply that in CCD0 we have found one way to simplify the cluster operator so as to at least substantially remove the failures of the cluster operator of CCD in the strongly correlated limit.  By working directly at the level of modifying the cluster operator, CCD0 ensures that we can follow all the usual machinery of standard coupled cluster theory straightforwardly.  In particular, we have a well-defined fermionic wave function.  Just as with standard CCD, we can define a left-hand wave function $\langle 0| (1+Z) \, \exp(-T)$ where $Z$ is a de-excitation operator which has, in CCD0, symmetric coefficients $z^{ij}_{ab}$ and with the aid of which we can construct density matrices and take expectation values.  We can in principle define excited states via equation of motion coupled cluster, and so on.

Of course the pieces of the cluster operator that CCD0 removes must be put back in somehow.  That is, we do not advocate CCD0 as a stand-alone method for the description of electronic systems, because the same-spin correlations have been removed from the theory entirely and results in the absence of strong correlation are poorer than what we could get from CCD itself.  Moreover, the opposite-spin correlation in CCD0 is not the same as in CCD either, as can be seen in the Hubbard Hamiltonian where there is no same-spin interaction, yet CCD and CCD0 deliver different results nonetheless.  One must, at the end of the day, correct CCD0 for what it has taken out of CCD.

Let this not, however, obscure the point of this paper: to the extent that one can identify which terms in the coupled cluster equations are responsible for the failures of the method, the failures can be resolved cheaply by simply removing those terms in a suitable way, or more expensively by adding additional terms which counteract the effect.  The former approach is computationally simpler and, we feel, may merit more attention than it has thus far received.  Identifying the culprit at the CCD level may help us design a selective introduction of higher cluster operators to balance their defects.  Work along these lines will be presented in due time.

By virtue of its exponential parameterization, single-reference coupled cluster theory can in principle deliver the correct coefficients on several important determinants without breaking down entirely, if we choose the correlator correctly; in other words, while based on a single reference determinant, the wave function can be multiconfigurational in nature.  It does not possess full flexibility to choose the coefficients on each determinant independently, but this is apparently not necessary to describe static correlation in many situations.  The successes of coupled cluster methods which describe strongly correlated systems by removing pieces of the cluster operator provide a fresh perspective on static correlation.  Coupled cluster methods which are based on a symmetry-adapted reference which is unstable toward broken symmetry can at least partially account for static correlation effects whenever the coupled-cluster energy is stable and well-behaved.

\begin{acknowledgments}
This work was supported by the National Science Foundation (CHE-1462434).  GES is a Welch Foundation chair (C-0036).  We would like to thank Carlos Jim\'enez-Hoyos for providing the full configuration interaction results for H$_4$.
\end{acknowledgments}

\bibliography{CCD0}

\begin{thebibliography}{44}
\bibitem [{Foo()}]{Footnote1}%
  \BibitemOpen
  \href@noop {} {}\bibinfo {note} {Here, "lm" denotes ladders + mosaics, which
  denotes the terms we keep in the CCD equations}\BibitemShut {NoStop}%
\end{thebibliography}

\end{document}